\documentclass[12pt]{article}
\input epsf
\usepackage{psfrag}
\usepackage{graphicx}

\font\mybb=msbm10 at 12pt
\def\bb#1{\hbox{\mybb#1}}

\renewcommand{\a}{\alpha}

\newcommand{\rmd}{{\rm d}}
\newcommand{\m}{\mu}
\newcommand{\n}{\nu}

\def\be{\begin{equation}}
\def\ee{\end{equation}}
\def\bea{\begin{eqnarray}}
\def\eea{\end{eqnarray}}
\def\ba{\begin{array}}
\def\ea{\end{array}}
\def\bi{\begin{itemize}}
\def\ei{\end{itemize}}

\catcode`\@=11
\def\@citex[#1]#2{%
\if@filesw \immediate \write \@auxout {\string \citation {#2}}\fi
\@tempcntb\m@ne \let\@h@ld\relax \def\@citea{}%
\@cite{%
  \@for \@citeb:=#2\do {%
    \@ifundefined {b@\@citeb}%
      {\@h@ld\@citea\@tempcntb\m@ne{\bf ?}%
      \@warning {Citation `\@citeb ' on page \thepage \space undefined}}%
      {\@tempcnta\@tempcntb \advance\@tempcnta\@ne%
      \@tempcntb\number\csname b@\@citeb \endcsname \relax%
      \ifnum\@tempcnta=\@tempcntb 
        \ifx\@h@ld\relax%
          \edef \@h@ld{\@citea\csname b@\@citeb\endcsname}%
        \else%
          \edef\@h@ld{\ifmmode{-}\else--\fi\csname b@\@citeb\endcsname}%
        \fi%
      \else
        \@h@ld\@citea\csname b@\@citeb \endcsname%
        \let\@h@ld\relax%
      \fi}%
    \def\@citea{,\penalty\@highpenalty\,}%
  }\@h@ld
}{#1}}

\def\@citeb#1#2{{[#1]\if@tempswa , #2\fi}}
\def\@citeu#1#2{{$^{#1}$\if@tempswa , #2\fi }}
\def\@citep#1#2{{#1\if@tempswa , #2\fi}}

\def\bcites{         
        \catcode`\@=11
        \let\@cite=\@citeb
        \catcode`\@=12
}

\def\upcites{         
        \catcode`\@=11
        \let\@cite=\@citeu
        \catcode`\@=12
}

\def\plaincites{      
        \catcode`\@=11
        \let\@cite=\@citep
        \catcode`\@=12
}

\newcount\hour
\newcount\minute
\newtoks\amorpm
\hour=\time\divide\hour by 60
\minute=\time{\multiply\hour by 60 \global\advance\minute by-\hour}
\edef\standardtime{{\ifnum\hour<12 \global\amorpm={am}%
        \else\global\amorpm={pm}\advance\hour by-12 \fi
        \ifnum\hour=0 \hour=12 \fi
        \number\hour:\ifnum\minute<10 0\fi\number\minute\the\amorpm}}
\edef\militarytime{\number\hour:\ifnum\minute<10 0\fi\number\minute}

\def\draftlabel#1{{\@bsphack\if@filesw {\let\thepage\relax
   \xdef\@gtempa{\write\@auxout{\string
      \newlabel{#1}{{\@currentlabel}{\thepage}}}}}\@gtempa
   \if@nobreak \ifvmode\nobreak\fi\fi\fi\@esphack}
        \gdef\@eqnlabel{#1}}
\def\@eqnlabel{}
\def\@vacuum{}
\def\marginnote#1{}
\def\draftmarginnote#1{\marginpar{\raggedright\scriptsize\tt#1}}
\def\draft{
        \pagestyle{plain}
        \overfullrule=two-point
        \oddsidemargin -.5truein
        \def\@oddhead{\sl \phantom{\today\quad\militarytime} \hfil
        \smash{\Large\sl DRAFT} \hfil \today\quad\militarytime}
        \let\@evenhead\@oddhead
        \let\label=\draftlabel
        \let\marginnote=\draftmarginnote
        \def\ps@empty{\let\@mkboth\@gobbletwo
        \def\@oddfoot{\hfil \smash{\Large\sl DRAFT} \hfil}
        \let\@evenfoot\@oddhead}
        \def\@eqnnum{(\theequation)\rlap{\kern\marginparsep\tt\@eqnlabel}%
        \global\let\@eqnlabel\@vacuum}  }

\begin{document}

\hfill CERN-TH/2002-012

\hfill {\tt hep-th/0201258} 
\vspace{-0.2cm}

\begin{center}
\Large
{ \bf Boundary multi-trace deformations and OPEs in AdS/CFT
  correspondence} 
\normalsize

\vspace{2cm}
\large
{\bf
Anastasios C. Petkou}
\footnote{tassos.petkou@cern.ch} \\
\normalsize

CERN Theory Division, \\
CH-1211 Geneva 23, \\
Switzerland
\vspace{.5cm}

\end{center}

\vspace{0.8cm}
\large
\centerline{\bf Abstract}
\normalsize
\vspace{.5cm}

We argue that multi-trace deformations of the boundary CFT in AdS/CFT
correspondence can arise
through the OPE of single-trace 
operators. We work out the example  of
a scalar field 
in AdS$_5$ with cubic self interaction. By an appropriate
reparametrization of the
boundary data we are able to deform the boundary CFT by
a marginal operator that couples to the conformal
anomaly. Our method can be used in the analysis  of
multi-trace deformations in ${\cal N}=4$ SYM where the OPEs of various
single-trace operators are known.

\vfill

\section{Introduction}

The AdS/CFT
correspondence \cite{adscft} is a concrete realization of the broad class of
holographic ideas. An implication of such ideas is that the 
classical equations of motion of fields that live in a curved bulk
space  determine the 
quantum states of a conformal field theory (CFT) that lives in the
boundary. In the prototype example of the AdS/CFT correspondence,
the equations of motion of classical IIB supergavity on AdS$_5\times$
S$^5$ determine the correlation functions of ${\cal N}=4$ SYM on $
  \bb{R}^4$. 

A remarkable success of the above example of AdS$_5$/CFT$_4$
correspondence is  the equivalence of  (part of)
the bulk and boundary spectra \cite{ferrara1}. For example, the 
Kaluza-Klein modes of IIB SUGRA on AdS$_5\times$ S$^5$ are in one-to-one
correspondence with 
chiral gauge invariant operators  of ${\cal N}=4$ SYM, which are realised as 
single-traces of its elementary fields.
However, it is also clear
that the boundary CFT spectrum contains an infinite number of
operators that do not correspond 
directly to any SUGRA modes. One class of such operators are the so
called Konishi-like operators that may be defined as classically conserved
currents of the non-interacting boundary CFT \cite{konishi}. These operators
correspond to massive string modes and hence cannot be seen in the
SUGRA approximation of IIB string theory. Another class of boundary
operators with no obvious SUGRA counterpart are multi-trace gauge
invariant operators of ${\cal N}=4$ SYM. They can be protected 
or non-protected by superconformal invariance \cite{ferrara}. After
some initial confusion regarding 
their status (see e.g. \cite{freedmist}), it was realized that
such operators should have a SUGRA realization since they arise in the
operator 
product expansions (OPEs) of the boundary operators as strong coupling
\cite{AFP1,AFP2}. The 
general perception is that they correspond to ``multi-particle''
supergravity  states
\cite{dhoker},
nevertheless we feel that their status is not yet fully clarified.

The study of
boundary multi-trace operators can be accomplished
{\it indirectly} through 
the operator product expansion of single-trace
operators. Such studies have lead to a number of non-trivial checks for
AdS/CFT \cite{AFP1,AFP2,dhoker,tests}, as well as to some unexpected new properties
\cite{EPSS}. One of them is the 
existence of operators whose scaling dimensions are
non-renormalized despite the fact that they are not in the BPS list
\cite{AFP1,AFP2}. A new
protection mechanism had to be invoked to  explain this phenomenon
\cite{AES}. Also, it was 
demonstrated that 
multi-trace operators can acquire finite non-zero anomalous 
dimensions at strong coupling, but the detailed mechanism of such a 
phenomenon   remains
unclear.

Recently the interest on multi-trace operators has
been revived due to ideas coming from string theory. Namely, it
has been argued that the ``multi-particle'' SUGRA states might
correspond to the 
``remnants'' of non-local string couplings \cite{ABS}.
Such ideas give rise to the practical question of how
explicit calculations of correlators involving multi-trace operators
can be done in the 
context of AdS/CFT. This question is in direct relevance to the
study of multi-trace deformations of the boundary
CFT. An answer to this question came with 
the works of 
\cite{BSS} and \cite{witten}, where it has been argued that
multi-trace deformations of the boundary CFT can by studied via
AdS/CFT by a generalization of the boundary conditions. Further
refinement was proposed in \cite{mueck} and \cite{minces} such that
both regular and irregular boundary conditions are properly taken into
account.

In this letter we approach the problem of
incorporating multi-trace deformations, directly from the boundary
CFT side. For non-zero 
sources the standard
AdS/CFT correspondence may be viewed as a  prescription for
deforming the 
strongly coupled boundary  CFT by single-trace operators coupled to
$x$-dependent couplings. When the boundary effective action is evaluated 
at least up to quartic order in the sources, it is possible
that the operators which appear in the OPE of the single-traces might
contribute to the deformation of the boundary CFT. In explicit AdS/CFT
examples such operators 
in general include a number of 
multi-trace operators which may be relevant, marginal or
irrelevant. Only the relevant and the marginal ones survive in the UV
limit. To illustrate our approach, we study here a simple example of a
scalar field in AdS$_5$ with cubic 
self interaction \cite{HPR}. By an appropriate
reparametrization of the $x$-dependent couplings (sources) we arrange so
that the boundary  CFT is perturbed by marginal operators. Moreover,
we  show that
the boundary conformal anomaly plays the role of the bare coupling for
such marginal deformations. Our methods can  be applied to
study the 
double-trace deformation in examples of AdS/CFT 
where the boundary four-point functions and OPEs are known, such as  ${\cal
  N}=4$ SYM in $d=4$ or the (2,0) tensor multiplet in $d=6$
\footnote{Of course, 
  there is no notion of single- or multi-trace operators  in
  the (2,0) tensor multiplet, nevertheless one can still apply OPE
  techniques to study the operator content of the theory at strong
  coupling \cite{gleb}.}
   \cite{gleb}.
\vfill

\section{Multi-trace deformations and boundary conditions}

The standard procedure for calculating correlations functions in a
boundary CFT from AdS/CFT correspondence may be schematically written as 
\be
\label{adscft}
Z_{SUGRA}[\phi_0]=\int({\cal D}\phi,\phi_0)e^{-S_{gr}[\phi]} \equiv
  e^{W_{R}[\phi_0]} \equiv \langle e^{\int \phi_0{\cal
      O}}\rangle_{CFT}\,.
\ee
In words this reads that performing the SUGRA path integral with
prescribed boundary conditions $\phi_0$ on the fields $\phi$ yields
the generating functional (effective action) $W_{R}[\phi_0]$ of
  connected renormalized correlation 
functions of the operator ${\cal O}$ in the boundary CFT. The last
  equality in (\ref{adscft}) shows that one may also view the AdS/CFT
  correspondence 
  as a definite prescription to deform the boundary CFT by the
  operator ${\cal O}$ coupled to the $x$-dependent coupling
  $\phi_0$. This latter 
interpretation may not be widely known, nevertheless there exist
  extensive studies 
of CFT deformations by $x$-dependent couplings \cite{hugh} and their
  general results can be applied to 
AdS/CFT.

The procedure described by (\ref{adscft}) shows the difficulties
with multi-trace deformations; since in explicit SUGRA calculations 
the fields $\phi$ are sources for  single-trace boundary operators one does 
not know how 
to calculate directly correlation functions of multi-trace operators,
or equivalently how to perturb the boundary CFT by multi-trace
deformations. A procedure to incorporate multi-trace operators has
been recently proposed in  \cite{BSS,witten}. The solution
to the second 
order bulk equations of
motion\footnote{We use throughout the Euclidean version of the
  Poincar\'e  
  patch of AdS$_{d+1}$
  where $\rmd x^{\m}\rmd x_{\m} = \frac{1}{r^2}(\rmd r^2 +\rmd
  \bar{x}^2)$ with $x=(r,\bar{x})$.}
for the SUGRA field $\phi$ behaves in general 
near the AdS$_{d+1}$ conformal boundary $r=0$ as  
\be
\label{BC1}
\phi(r,\bar{x})|_{r\rightarrow
  0}=r^{d-\Delta}[\phi_{0}(\bar{x})+O(r^2)]+
  r^{\Delta}[A(\bar{x})+O(r^2)]\,,\,\,\,\,\, \Delta
  =\frac{d}{2}+\frac{1}{2}\sqrt{d^2+4m^2}\,,
\ee
where $m^2\geq-\frac{d^2}{4}$ is the AdS mass.
The parameter $\Delta$ becomes the scaling dimension of the boundary
  operator ${\cal O}$.\footnote{We assume here that
  $\Delta\geq\frac{d}{2}$ as is relevant for our model as well as for
  the the standard IIB/${\cal N}=4$ SYM duality.} 
To calculate now the expectation value  $\langle{\rm exp}[\int \phi_0{\cal
  O}]\rangle$ in the boundary CFT one needs the leading behaviour of
  $\phi(r,\bar{x})$ near the 
  boundary which is
  given by the first term in 
  (\ref{BC1}). Nevertheless, since it has been shown \cite{klebanov}
  that $A(\bar{x})$ 
  in (\ref{BC1}) corresponds to the expectation value of
  the operator ${\cal O}$, the result of the expectation value
  calculation above 
  can be schematically represented as ${\rm exp}[\int \phi_0
  A]$. Now, the proposal of \cite{witten} is that if one wants
  to calculate the expectation value  $\langle {\rm exp}[\int
  \hat{\phi}_0{\cal O} +
  \int{\cal F}({\cal O},\rmd{\cal O})]\rangle $, where ${\cal F}({\cal
  O},\rmd{\cal O})$ is a generic functional of the operator ${\cal
  O}$, then one should solve the same bulk equations as the ones
  leading to (\ref{BC1}) but with modified  boundary conditions such
  that 
\be
\label{BC2}
\phi_0(\bar{x})= \hat{\phi}_0(\bar{x}) + \frac{\delta{\cal F}(A,\rmd
  A)}{\delta A(\bar{x})}\,.
\ee
Then, $\hat{\phi}_0$ in (\ref{BC2}) provides the source to the
boundary operator ${\cal O}$ 
  and the remaining term the source to the multi-trace operator represented by
  the functional ${\cal F}({\cal O},\rmd{\cal O})$. Refinements of
  the above proposal to include a proper treatment of both the so
  called  
  regular and irregular boundary modes were presented in
  \cite{mueck,minces}. 

\section{The boundary action to quartic order}

The proposal for incorporating multi-trace deformations of the
boundary CFT by modifying the boundary conditions can in principle be
used to perform explicit calculations in specific AdS/CFT
models. As we mentioned in the Introduction, an alternative way of
incorporating multi-trace operators in 
the boundary effective action could be through the OPE of single-trace
operators. This entails the calculation of  the
boundary effective action up to quartic order in the sources to read
off the four-point functions of the single-trace operators. One
advantage of this approach is, nevertheless, that one can use the
known results for the 
four-point functions of some boundary CFTs from AdS/CFT
correspondence, such as  
${\cal N}=4$ SYM to study their
corresponding boundary deformations. The purpose of this work is to
illustrate our method in the toy 
model of a scalar field in AdS$_5$ with cubic self-interaction
\cite{HPR}.
The classical massive action is
\be
\label{action}
S_{gr}[\phi] = \frac{1}{2}\int \rmd^5x\sqrt{g}(g^{\m\n}\partial_\m \phi
\partial_\n\phi +m^2 \phi^2)
+\frac{\lambda}{3!}\int\rmd^5x\sqrt{g}\phi^3\,,
\ee
where $\lambda$ is an AdS coupling constant. The equations of motion are
\be
\label{eom}
(\nabla^2-m^2)\phi =\frac{\lambda}{2}\phi^2\,.
\ee
One needs to solve the equations of motion (\ref{eom}) subject to
boundary conditions imposed at $r=0$ and substitute their solution back
to (\ref{action}). This way one obtains a functional of the
boundary conditions which is interpreted as the generating functional
for connected renormalized correlation functions of the boundary
CFT. A nice way 
to accomplish that  is to first solve an intermediate problem
which amounts to exactly the same procedure as described above but with the
boundary conditions now imposed on some hypersurface of AdS$_5$ {\it
  near} the boundary as
\be
\label{DBC}
\phi(r,\bar{x})|_{r=\epsilon} = \phi_{\epsilon}(\bar{x})\,,\,\,\,\,\,
\epsilon <<1\,.
\ee
Solving now the equations of motion (\ref{eom}) with the boundary
condition (\ref{DBC}) and substituting their solution back into
(\ref{action}) we will 
get a functional of $\phi_{\epsilon}$. The latter is then interpreted
as the {\it regularized} generating functional of the boundary CFT. In doing
the above, one must always keep in mind that the $r$ integration
should also be restricted to the range $r\in [\epsilon,\infty)$. 

The procedure above has been described in a number of works and here
we recapitulate its essential points. The general solution of the
non-homogeneous equation of motion (\ref{eom}) with the boundary condition
(\ref{DBC}) is
\be
\label{soleom}
\phi(r,\bar{x}) = \bar{\phi}(r,\bar{x}) +
\frac{\lambda}{2}\int\rmd^5y\sqrt{g} G_{\epsilon}(x,y)\phi^{2}(y)\,,
\ee
where $\bar{\phi}(r,\bar{x})$ is a solution of the homogeneous part of
(\ref{eom}) that satisfies the boundary condition (\ref{DBC}) and
$G_{\epsilon}(x,y)$ is a Green's function of the homogeneous part of
(\ref{eom}) that vanishes when either of its arguments lies on the
``boundary'' $r=\epsilon$. The latter can be written as
\be
\label{greenf}
G_{\epsilon}(x,y) = G(x,y) + F(x,y)\,,
\ee
where 
\be
\label{greenf0}
(\nabla^2 -m^2)G(x,y) = \delta^{d}(x-y)\,,\,\,\,\,\, (\nabla^2
  -m^2)F(x,y) = 0\,,\,\,\,\,\, F(x,y)|_{x,y\in
  \partial_{\epsilon}} = -G(x,y)|_{x,y\in\partial_{\epsilon}}\,,
\ee
and $\partial_{\epsilon}$ denotes the ``boundary'' surface
  $r=\epsilon$ of AdS$_5$.
Explicit expression for the above quantities exist since the early
days of AdS/CFT \cite{2ptfunc} and although they are not needed for
our calculations 
here we give them for completeness in the Appendix. We only present here the
so called bulk-to-boundary propagator that allows one to reconstruct
the bulk field given the boundary condition (\ref{DBC}) as
\bea
\label{barphi}
\bar{\phi}(r,\bar{x}) &=& \int\rmd^4
\bar{x}'K(r;\bar{x},\epsilon;\bar{x}') \phi_{\epsilon}(\bar{x}')\,,\\
\label{K}
K(r;\bar{x},\epsilon;\bar{x}')&=&
\left(\frac{r}{\epsilon}\right)^{2}
\int\frac{\rmd^4\bar{p}}{(2\pi)^4} e^{-{\rm
    i}\bar{p}(\bar{x}-\bar{x}')} \frac{{\rm K}_{\a}(|\bar{p}|r)}{{\rm
    K}_{\a}(|\bar{p}|\epsilon)}\,, \,\,\,\,\a=\Delta-2\,,
\eea
where ${\rm K}_{\a}(z)$ are the standard modified Bessel functions.
It is important to note that in writing (\ref{barphi}) and
(\ref{K}) we tacitly imposed the condition that the bulk scalar field
vanishes as $r\rightarrow \infty$. Substituting the above into the
action (\ref{action}) one obtains up to quartic  order in the
boundary condition
\bea
\label{Se}
S[\phi_{\epsilon}] &=& -\frac{1}{2}\int\rmd^4\bar{x}\,\epsilon^{-3}
\phi_{\epsilon}(\bar{x}) \partial_r\phi(r,\bar{x})|_{r=\epsilon}
\nonumber \\
&&\hspace{-.7cm}+\frac{\lambda}{3!}\int\rmd^4\bar{x}_1\rmd^4\bar{x}_2
\rmd^4\bar{x}_3\,\phi_{\epsilon}(\bar{x}_1)\phi_{\epsilon}(\bar{x}_2)
\phi_{\epsilon}(\bar{x}_3)\,\Pi_{3,\epsilon}(  
\bar{x}_1,\bar{x}_2,\bar{x}_3) \nonumber \\
&&\hspace{-.7cm}
+ \frac{\lambda^2}{8}\int\rmd^4\bar{x}_1\rmd^4\bar{x}_2\rmd^4\bar{x}_3 
\rmd^4\bar{x}_4\,\phi_{\epsilon}(\bar{x}_1)\phi_{\epsilon}(\bar{x}_2)
 \phi_{\epsilon}(\bar{x}_3)
\phi_{\epsilon}(\bar{x}_4)\,\Pi_{4,\epsilon}(  
\bar{x}_1,\bar{x}_2,\bar{x}_3,\bar{x}_4)\,,\\ 
\label{P3}
\Pi_{3,\epsilon}(\bar{x}_1,\bar{x}_2,\bar{x}_3) &=&\int_{\epsilon}^{\infty}\rmd
  rr^{-5}\int\rmd^4{\bar{x}}
  K(r;\bar{x},\epsilon;\bar{x}_1) K(r;\bar{x},\epsilon;\bar{x}_2)
  K(r;\bar{x},\epsilon;\bar{x}_3) \,,
  \\
\label{P4}
\Pi_{4,\epsilon}(\bar{x}_1,\bar{x}_2,\bar{x}_3, \bar{x}_4)
  &=&\int_{\epsilon}^{\infty}\rmd  
  r\rmd r'(rr')^{-5}\int\rmd^4\bar{x}\rmd^4\bar{y}\Bigl[
  K(r;\bar{x},\epsilon;\bar{x}_1) K(r;\bar{x},\epsilon;\bar{x}_2)
  G_{\epsilon}(x,y) \nonumber \\
 &&\hspace{5cm}K(r';\bar{y},\epsilon;\bar{x}_3)
  K(r';\bar{y},\epsilon;\bar{x}_4)\Bigl]\,. 
\eea
From (\ref{barphi}) and (\ref{K}) one easily obtains the asymptotic
behaviour of $\phi_{\epsilon}$ near $\epsilon
\rightarrow 0$ as
\be
\label{phie0}
\phi_{\epsilon}(\bar{x})|_{\epsilon\rightarrow
  0}=\epsilon^{4-\Delta}[\phi_0(\bar{x})+O(\epsilon^2)]
  +\epsilon^{\Delta}[A(\bar{x})+O(\epsilon^2)] \,.
\ee 
The functions $\phi_0(\bar{x})$ and $A(\bar{x})$ would be  two
  independent boundary data necessary for the complete solution of the
  second order bulk equation of motion (\ref{eom}), nevertheless due to the
  imposed 
  regularity of the bulk solution implied by (\ref{K}) 
  there is a relationship between them as
\be
\label{phiA}
A(\bar{x}) = C_{\Delta}\int\rmd^4
\bar{x}'\frac{\phi_0(\bar{x}')}{(\bar{x}-\bar{x}')^{2\Delta}}\,,\,\,\,\,\,
C_{\Delta}= \frac{\Gamma(\Delta)}{\pi^{2}\Gamma(\Delta
  -2)}\,. 
\ee
For $r=0$ the boundary
  action is a functional of the boundary data $\phi_0$. 
For example, to quadratic order in $\phi_0$ one finds
\bea
\label{Stwo-point}
S_{\epsilon}[\phi_0] &=& \frac{1}{2}\int
\rmd^4\bar{x}_1\rmd^4\bar{x}_2\phi_0(\bar{x}_1)
\phi_0(\bar{x}_2)\Pi({\bar{x}_{12}},\epsilon)\,,\\
\label{two-point}
\Pi({\bar{x}_{12}},\epsilon)&=&
\epsilon^{4-2\Delta}\int\frac{\rmd^4\bar{p}}{(2\pi)^4} e^{-{\rm
    i}\bar{p}\bar{x}_{12}}\left[(4-\Delta)-|\bar{p}|\epsilon\frac{
    {\rm K}_{\a-1}(|\bar{p}|\epsilon)}{{\rm K}_\a(|\bar{p}|\epsilon)}\right]\,.
\eea
The action (\ref{Stwo-point}) is interpreted as the regularized
action of the boundary CFT since in general it contains
a finite number of divergent terms as $\epsilon\rightarrow 0$. One can
subtract these divergences by introducing local counterterms
built out from the field $\phi(r,\bar{x})$ at $r=\epsilon$. This 
subtraction procedure amounts to a specific choice of renormalization
scheme as the counterterms include finite parts \cite{kostas}. After
subtraction, 
one can take the limit $\epsilon\rightarrow 0$ and is left with an
action that is the generating 
functional $W_R[\phi_0]$ of connected renormalized correlation functions of the
operator ${\cal O}$. However, in order to apply the OPE we need the
{\it partition function} of the boundary CFT which  is the generating
functional of {\it both} the connected and disconnected $n$-point
functions. Up to four-point functions this reads
\bea
\label{Sren}
Z_R[\phi_0] \equiv e^{W_R[\phi_0]}&=& 1+\frac{1}{2}\int
\rmd^4\bar{x}_1\rmd^4\bar{x}_2\,\phi_0(\bar{x}_1)
\phi_0(\bar{x}_2)\,\langle{\cal 
  O}(\bar{x}_1) {\cal
  O}(\bar{x}_2) \rangle_R\nonumber \\
&&
\hspace{-2.5cm}-\frac{\lambda}{3!}\int\rmd^4\bar{x}_1\rmd^4\bar{x}_2
\rmd^4\bar{x}_3\,\phi_0(\bar{x}_1)\phi_0(\bar{x}_2)\phi_0(\bar{x}_3)
\,\langle{\cal   
 O}(\bar{x}_1) {\cal O}(\bar{x}_2) {\cal
  O}(\bar{x}_3) \rangle_R\nonumber \\
&& \hspace{-2.5cm} +\frac{\lambda^2}{4!}\int\rmd^4\bar{x}_1\rmd^4\bar{x}_2
\rmd^4\bar{x}_3\rmd^4\bar{x}_4
\,\phi_0(\bar{x_1})\phi_0(\bar{x}_2)\phi_0(\bar{x}_3)\phi_0(\bar{x}_4)  
\,\langle{\cal   
 O}(\bar{x}_1) {\cal O}(\bar{x}_2) {\cal
  O}(\bar{x}_3) {\cal O}(\bar{x}_4) \rangle_R\,,
\eea
where the symmetry factors have been changed according to the
requirement that the correlation functions are totally symmetric with
respect to permutations of their arguments. For completeness, the
explicit formulae for the various correlators in (\ref{Sren}) are
given in the Appendix and we refer the reader to \cite{HPR} for the
calculations of the integrals.\footnote{Despite that fact
  that a complete analysis of the counterterms needed to
renormalized the $\epsilon\rightarrow 0$
singularities in $n$-point functions for $n\geq 3$ has not been
performed, one can obtain explicit expressions for non-coincident
arguments by simply taking the $\epsilon\rightarrow 0$ limit of
(\ref{K}), (\ref{P3}) and (\ref{P4}) before performing the integrals.}
 
If one has managed to remove completely all scale dependence from
(\ref{Sren}), then one would find  the generating functional of a
quantum CFT.\footnote{Notice that logarithms of the invariant ratios
  in four-point functions 
  are in perfect agreement with conformal invariance since they do not
  need an arbitrary mass scale for their proper definition.} This
is in general possible, however for special values 
of the dimension $\Delta$ one cannot do this. As it was shown in
\cite{PS} following the ideas of \cite{OP}, for integer values of the
parameter $\a$ the renormalized correlation functions in (\ref{Sren})
necessarily contain logarithms which break conformal
invariance. The relevant case for our present work is when 
\be
\label{dimanom}
\Delta = 2+k\,,\,\,\, k=0,1,2,..\,.
\ee
Then, the renormalized  two-point function 
\be
\langle{\cal
  O}(\bar{x}_1) {\cal
  O}(\bar{x}_2) \rangle_R=C_{\cal
  O}\left[\frac{1}{\bar{x}_{12}^{2\Delta}}\right]_{R} \,,
\ee
contains logarithms and this implies the existence of a
conformal anomaly in the theory as \cite{PS}
\be
\label{anom}
\langle\Theta\rangle\equiv \int\rmd^4\bar{x}\langle
  T_{\m\m}(\bar{x})\rangle = \frac{1}{2}{\cal 
  P}_k\int\rmd^4\bar{x}\,\partial^{2k} \,\phi^2_{0}(\bar{x})\,, \,\,\,\,{\cal
  P}_k = C_{\cal O}\frac{2\pi^{2}}{2^{2k}\Gamma(k+1)\Gamma(k+2)}\,.
\ee
For example, when $\Delta=2$ using the
  correct normalization for the two-point function $C_{\cal
  O}=1/2\pi^2$ we have 
\be
\label{anom2}
\langle\Theta\rangle = \frac{1}{2}\int\rmd^4\bar{x}\,\phi_0^2(\bar{x})\,.
\ee
In this case only the two-point function contributes to the conformal
anomaly.

\section{Operator deformations and the OPE}

The essential observation now is that once the renormalized effective
action of the boundary CFT is calculated up to quartic order in the
sources, operators that appear in the OPE of two ${\cal O}$'s are 
naturally incorporated into it. In the case of ${\cal N}=4$ SYM such
operators will in general include both single- as 
well as  
double-traces. In the simple model (\ref{action}), the OPE analysis of
the four-point  
function in (\ref{Sren}) has been done in whole generality in
\cite{HPR} and here we summarise the essential points. The OPE of the
fields ${\cal O}$ that reproduces the explicit form of the
four-point function in (\ref{Sren}) 
can be schematically written as
\be
\label{OPE}
{\cal O}(\bar{x}_1){\cal O}(\bar{x}_2) =\langle{\cal
  O}(\bar{x}_1){\cal O}(\bar{x}_2)\rangle_R +
\sum_{\{O\}}\frac{G_{O}}{C_O}\,\frac{
  1}{(\bar{x}^2_{12})^{\Delta-\frac{\Delta_O}{2}}}{\cal
  C}(\bar{x}_{12}, \partial_{\bar{x}_2})\cdot O(\bar{x}_2)\,,
\ee
where $\{O\}$ is an infinite set of operators with corresponding dimensions
$\Delta_0$ and even spin, while the dot in (\ref{OPE}) denotes the
appropriate tensor contraction. The OPE coefficients ${\cal
  C}(\bar{x}_{12},\partial_{\bar{x}_2})$ are complicated non-local
expression which are explicitly known. For example, when $O$ is a
scalar we have 
\bea
\label{OPEcoef}
{\cal C}_{O}(\bar{x}_{12},\partial_{\bar{x}_2}) &=&
\frac{1}{B(\frac{\Delta}{2}, \frac{\Delta}{2})} \!\int_0^1\!\!\rmd
t[t(1-t)]^{\frac{\Delta}{2}-1} \sum_{m=0}^{\infty}
\frac{(-1)^m\Gamma(\Delta+1-\frac{d}{2})}{m!\Gamma(\Delta+1+m-\frac{d}{2})}
\left[\frac{\bar{x}^2_{12}}{4}t(1-t)\right]^m
\!\!\!\partial_{\bar{x}_2}^{2m}e^{t\bar{x}_{12}\cdot 
 \partial_{\bar{x}_2}}\nonumber \\
&=& 1+\frac{1}{2}(\bar{x}_{12})_\m\partial_{\bar{x}_2,\m}
+\frac{\Delta +2}{8(\Delta+1)}
(\bar{x}_{12})_\m(\bar{x}_{12})_\m
\partial_{\bar{x}_2,\m}\partial_{\bar{x}_2,\n}\nonumber \\
&&
-\frac{\Delta}{16(\Delta+1)(\Delta +1-\frac{d}{2})}
(\bar{x}^2_{12})\partial_{\bar{x}_2}^2 +\cdots  \,. 
\eea 
The parameter
$G_{O}$ is the unique coupling constant in the three-point function
$\langle O{\cal O}{\cal O}\rangle$ and $C_O$ is the normalization
constant of the two-point function $\langle OO\rangle$. The infinite set of
operators $\{O\}$ are classified according to increasing powers of
their dimension and spin. In our case, the first
few operators that contribute 
singular and marginal terms as $\bar{x}_{12}\rightarrow 0$ in
(\ref{OPE})   are the operator ${\cal O}$
itself, a scalar  operator with dimension $2\Delta+\lambda^2\gamma_*$
which we denote as 
${\cal O}^2$ and the energy momentum tensor which is a spin-2 operator
with dimension $d$. The quantity $\gamma_*$ is the anomalous dimension
of the operator ${\cal O}^2$ which can be explicitly calculated from
the results in \cite{HPR}. 
In general, an OPE such as (\ref{OPE}) is supposed to hold when the
two operators are close to each other. However, there is strong evidence
(see e.g. \cite{fradkin}) that in CFTs the OPE is an
analytic partial wave expansion, even for $D>2$, and as such it can be
inserted into the generating functional (\ref{Sren}) yielding, (taking
care of the combinatorics) 
\bea
\label{SrenOPE}
Z_R[\phi_0]&=& 1+\frac{1}{2}\int
\rmd^4\bar{x}_1\rmd^4\bar{x}_2\,\phi_0(\bar{x}_1)\phi_0(\bar{x}_2)\,
\langle{\cal 
  O}(\bar{x}_1) {\cal
  O}(\bar{x}_2) \rangle_R\nonumber \\
&&
\hspace{-2cm}-\frac{\lambda}{2}\int\rmd^4\bar{x}_1\rmd^4\bar{x}_2
\rmd^4\bar{x}_3\,\phi_0(\bar{x}_1)\phi_0(\bar{x}_2)\phi_0(\bar{x}_3)
\frac{G_{\cal O}}{C_{\cal O}}\frac{
  1}{(\bar{x}^2_{32})^{\frac{\Delta}{2}}}\,{\cal
  C}(\bar{x}_{32}, \partial_{\bar{x}_2})\cdot
\langle{\cal    
 O}(\bar{x}_2) {\cal O}(\bar{x}_1) \rangle_R\nonumber \\
&& \hspace{-2cm} +\frac{\lambda^2}{4}\int\rmd^4\bar{x}_1..\rmd^4\bar{x}_4
\,\phi_0(\bar{x}_1)..\phi_0(\bar{x}_4)  
\sum_{\{O\}}\frac{G_{O}}{C_O}\frac{
  1}{(\bar{x}^2_{43})^{\Delta-\frac{\Delta_O}{2}}}\,{\cal
  C}(\bar{x}_{43}, \partial_{\bar{x}_3})\cdot \langle O(\bar{x}_3){\cal   
 O}(\bar{x}_1) {\cal O}(\bar{x}_2)\rangle_R\nonumber \\
&&\hspace{-2cm} +\frac{1}{8}\int\rmd^4\bar{x}_1..\rmd^4\bar{x}_4
\,\phi_0(\bar{x}_1)..\phi_0(\bar{x}_4) \langle{\cal 
  O}(\bar{x}_1) {\cal
  O}(\bar{x}_2) \rangle_R \langle{\cal 
  O}(\bar{x}_3) {\cal
  O}(\bar{x}_4) \rangle_R\,.
\eea
From (\ref{SrenOPE}) we see that when we view the AdS/CFT
correspondence 
as a deformation of the boundary CFT with an
$x$-dependent coupling constant $\phi_0(\bar{x})$, 
 many operators enter naturally
into the partition function via the OPE. Of course, keeping the
$x$-dependent coupling  non-zero in general  breaks
the conformal invariance. The meaning then of equations such as
(\ref{SrenOPE}) is that conformal invariance is broken by perturbing
the CFT by all possible operators that appear in the OPE (\ref{OPE})
\cite{cardy}. 

Given the form of the generating
functional (\ref{SrenOPE}) it is possible to choose exactly which
operator deforms the boundary CFT by appropriately  adjusting
the sources $\phi_0$. Such an adjustment of the sources $\phi_0$ has
a dual interpretation; from the point of view of the bulk theory 
it corresponds to a modification of  the boundary conditions
(\ref{phie0}) \cite{BSS,witten}, while from the point of view of the
boundary theory it corresponds to a reparametrization of the
$x$-dependent couplings. Namely, setting 
\be
\label{adj}
\phi_0(\bar{x}) = \hat{\phi}_0(\bar{x}) + \phi_1(\bar{x})\,,
\ee
we find from (\ref{SrenOPE})
\bea
\label{pert}
Z_R[\phi_0] &=& Z_R[\hat{\phi}_0] +\int\rmd^4\bar{x}_1\rmd^4\bar{x}_2
\, \phi_1(\bar{x}_1)\hat{\phi}_0(\bar{x}_2)\, \langle
{\cal O}(\bar{x}_1){\cal O}(\bar{x}_2)\rangle_R \nonumber \\
&& - \frac{\lambda}{2}
\int \rmd^4\bar{x}_1\rmd^4\bar{x}_2\rmd^4\bar{x}_3\,
\phi_1(\bar{x}_1)\hat{\phi}_0(\bar{x}_2)\hat{\phi}_0(\bar{x}_3) \, \langle
{\cal O}(\bar{x}_1){\cal O}(\bar{x}_2){\cal O}(\bar{x}_3)\rangle_R +\cdots\,.
\eea
To proceed now we have to take into account an important point that
follows from  the analysis of \cite{HPR}. As we have mentioned, the
four-point function in (\ref{Sren}) contains {\it both} connected and
disconnected parts. In the normalization of (\ref{Sren}) the
disconnected part comes with a factor $\lambda^{-2}$ in front. It
has been observed in \cite{HPR} that in generic AdS graphs the
contribution to the four-point function from the operator ${\cal O}$ itself
comes {\it only} from the connected part. This means that the
form of the OPE (\ref{OPE}) implies
\bea
\label{WRexp}
Z[\hat{\phi}_0] &=& 1+\frac{1}{2}\int
\rmd^4\bar{x}_1\rmd^4\bar{x}_2 \,\hat{\phi}_0(\bar{x}_1)\hat{\phi}_0(
\bar{x}_2)\,\langle{\cal 
  O}(\bar{x}_1) {\cal
  O}(\bar{x}_2) \rangle_R\nonumber \\
&&
\hspace{-1.5cm}-\frac{\lambda G_{\cal O}}{2C_{\Delta}}\int\rmd^4\bar{x}_2 
\rmd^4\bar{x}_3\,\hat{\phi}_0(\bar{x}_2)\hat{\phi}_0(\bar{x}_3)
\,\frac{
  1}{(\bar{x}^2_{32})^{\frac{\Delta}{2}}}\,{\cal 
  C}_{\cal O}(\bar{x}_{32}, \partial_{\bar{x}_2})A(\bar{x}_2)
\nonumber \\
&& \hspace{-1.5cm} +\frac{\lambda^2G_{\cal O}}{4C_{\cal
    O}}\int\rmd^4\bar{x}_1..\rmd^4\bar{x}_4\, 
\hat{\phi}_0(\bar{x}_1)..\hat{\phi}_0(\bar{x}_4)\,  
\frac{
  1}{(\bar{x}^2_{43})^{\frac{\Delta}{2}}}\,{\cal
  C}_{\cal O}(\bar{x}_{43}, \partial_{\bar{x}_3}) \langle {\cal
  O}(\bar{x}_3){\cal    
 O}(\bar{x}_1) {\cal O}(\bar{x}_2)\rangle_R \nonumber \\
&&\hspace{-1.5cm} +\frac{G_{{\cal O}^2}(\lambda)}{4C_{{\cal
    O}^2}(\lambda)}\int\rmd^4\bar{x}_1..\rmd^4\bar{x}_4 
\,\hat{\phi}_0(\bar{x}_1)..\hat{\phi}_0(\bar{x}_4)
\,(\bar{x}^2_{34})^{\frac{\lambda^2\gamma_*}{2}} \,{\cal
  C}_{{\cal O}^2}(\bar{x}_{43}, \partial_{\bar{x}_3}) \langle {\cal
  O}^2(\bar{x}_3){\cal    
 O}(\bar{x}_1) {\cal O}(\bar{x}_2)\rangle_R \nonumber \\
&&\hspace{-1.5cm} +\frac{1}{8}\int\rmd^4\bar{x}_1..\rmd^4\bar{x}_4 
\,\hat{\phi}_0(\bar{x}_1)..\hat{\phi}_0(\bar{x}_4) \,\langle{\cal 
  O}(\bar{x}_1) {\cal
  O}(\bar{x}_2) \rangle_R\langle{\cal 
  O}(\bar{x}_3) {\cal
  O}(\bar{x}_4) \rangle_R + \cdots \,,
\eea
where the dots represent terms involving higher orders in $\lambda$
and correlation functions involving tensor operators, while we have also
used (\ref{phiA}). 
The meaning of (\ref{WRexp}) is that the coupling and scaling dimension of
the operator ${\cal O}^2$ contain both $O(\lambda^0)$ as well as
$O(\lambda^2)$ contributions. This is generically true for all
operators in the OPE (\ref{OPE}) except ${\cal O}$ itself. We 
have seen already the $\lambda$-dependence for the scaling dimension of ${\cal 
  O}^2$, while for the coupling (more precisely, the ratio of the
coupling the the two-point function normalization),  we may write
\be
\label{couplexp}
\frac{G_{{\cal O}^2}(\lambda)}{C_{{\cal
    O}^2}(\lambda)}= \left.\frac{G_{{\cal O}^2}}{C_{{\cal
    O}^2}}\right|_0[1+\lambda^2 b_*+O(\lambda^3)]\,,
\ee
where the subscript 0 denotes the $\lambda$-independent part coming
    from the disconnected graphs in the four-point function. The
    finite number $b_*$ can be read from the results in \cite{HPR}.

We now observe  that if the two $\phi_1$-dependent terms
of (\ref{pert}) cancel the third and fourth terms in the rhs of
(\ref{WRexp}), then 
$Z_R[\hat{\phi}_0]$ would  generate
correlation functions  of ${\cal O}$ with ${\cal O}^2$
insertions. This can be arranged if the following integral equations
are satisfied
\be
\label{cond1}
\frac{C_{\cal O}}{C_{\Delta}} \int\rmd^4\,\bar{x}\phi_1(\bar{x})
A(\bar{x}) - \frac{\lambda G_{\cal O}}{2C_{\Delta}}\int\rmd^4\bar{x}_2 
\rmd^4\bar{x}_3\,\hat{\phi}_0(\bar{x}_2)\hat{\phi}_0(\bar{x}_3)
\,\frac{
  1}{(\bar{x}^2_{32})^{\frac{\Delta}{2}}}\,{\cal
  C}_{\cal O}(\bar{x}_{32}, \partial_{\bar{x}_2})A(\bar{x}_2)=0\,,
\ee
and
\bea
\label{cond2}
&&\frac{\lambda}{2}\int\rmd^4\bar{x}_1\rmd^4\bar{x}_2\rmd^4\bar{x}_3
\phi_1(\bar{x}_1) \,\hat{\phi}_0(\bar{x}_2)\hat{\phi}_0(\bar{x}_3)
\,\langle {\cal O}(\bar{x}_1){\cal O}(\bar{x}_2){\cal
  O}(\bar{x}_3)\rangle_R\nonumber \\
&&\hspace{0.8cm}- \frac{\lambda^2G_{\cal O}}{4C_{\cal
    O}}\int\rmd^4\bar{x}_1..\rmd^4\bar{x}_4 
\,\hat{\phi}_0(\bar{x}_1)..\hat{\phi}_0(\bar{x}_4)  
\frac{
  1}{(\bar{x}^2_{43})^{\frac{\Delta}{2}}}\,{\cal
  C}_{\cal O}(\bar{x}_{43}, \partial_{\bar{x}_3})\langle {\cal
  O}(\bar{x}_3){\cal    
 O}(\bar{x}_1) {\cal O}(\bar{x}_2)\rangle_R=0\,.
\eea
From (\ref{cond1}) we find 
\be
\label{phi1}
\phi_1(\bar{x}) = \frac{\lambda G_{\cal O}}{2C_{\cal O}}\int\rmd^4\bar{x}_2 
\rmd^4\bar{x}_3\,\hat{\phi}_0(\bar{x}_2)\hat{\phi}_0(\bar{x}_3)\,
\frac{
  1}{(\bar{x}^2_{32})^{\frac{\Delta}{2}}}\,{\cal
  C}_{\cal O}(\bar{x}_{32},
\partial_{\bar{x}_2})\delta^4(\bar{x}-\bar{x}_2)\,.
\ee
Substituting (\ref{phi1}) into (\ref{cond2}) and integrating the delta
function by parts we see that (\ref{cond2}) is also satisfied.
Therefore, by reparametrizing the initial sources as in (\ref{adj})
and (\ref{phi1}) one gets a new
partition function $Z_R[\hat{\phi}_0]$ that corresponds to a
deformation of the original CFT by the operator ${\cal O}^2$. For
example, the leading correction to the {\it connected} two-point
function would then read
\bea
\label{two-pointpert}
\langle{\cal O}(\bar{x}_1){\cal O}(\bar{x}_2)\rangle_{R}' &=&
\langle{\cal O}(\bar{x}_1){\cal O}(\bar{x}_2)\rangle_R \nonumber \\
&&\hspace{-2.5cm}+\left.\frac{G_{{\cal O}^2}}{2C_{{\cal
    O}^2}}\right|_0\int\rmd^4\bar{x}_3\rmd^4\bar{x}_4\, 
\hat{\phi}_0(\bar{x}_3)\hat{\phi}_0(\bar{x}_4)\,[1+\frac{\lambda^2\gamma_*}{2}
\ln(\bar{x}^2_{34}\mu^2)+\lambda^2b_*]\times \nonumber \\
&&\times \,{\cal
  C}_{{\cal O}^2}(\bar{x}_{43}, \partial_{\bar{x}_3}) \langle {\cal
  O}^2(\bar{x}_3){\cal    
 O}(\bar{x}_1) {\cal O}(\bar{x}_2)\rangle_R+\cdots\,,
\eea
where the arbitrary mass parameter $\mu$ is necessary for the correct
definition 
of the logarithm. The dots in (\ref{two-pointpert}) correspond to
terms of order $O(\lambda^4)$ as well as to three-point  functions 
involving scalar 
operators with dimensions greater that $2\Delta$ and operators
with non-zero (even) spin \cite{HPR}. Now, all angular dependent terms
in the integrand on the rhs of (\ref{two-pointpert}) drop out. This means that
one is left with three-point  functions involving only scalar
operators. Finally, we restrict ourselves to the case when $\Delta
=2$. Then, from translation invariance and renormalization
arguments one  obtains
\bea
\label{two-pointfinal}
\langle{\cal O}(\bar{x}_1){\cal O}(\bar{x}_2)\rangle_{R}' &=&
\langle{\cal O}(\bar{x}_1){\cal O}(\bar{x}_2)\rangle_R \nonumber \\
&&\hspace{-4cm}+\left.\frac{G_{{\cal O}^2}}{2C_{{\cal
    O}^2}}\right|_0\int\rmd^4\bar{x}_4 
\,\hat{\phi}_0^2(\bar{x}_4)\int \rmd^4\bar{x}_3\,
[1+\frac{\lambda^2\gamma_*}{2} 
\ln(\bar{x}^2_{3}\mu^2)+\lambda^2b_*] \,\langle {\cal
  O}^2(\bar{x}_3){\cal    
 O}(\bar{x}_1) {\cal O}(\bar{x}_2)\rangle_R+\cdots\,,
\eea
where now the dots in (\ref{two-pointfinal}) represent multiple ${\cal O}^2$
insertions coming from higher correlation functions. Insertions from
scalar operators with dimensions greater than $2\Delta =4$ are
irrelevant and drop out. One way to
see how (\ref{two-pointfinal}) is 
derived from (\ref{two-pointpert}) is to introduce a short-distance cut-off
$L\ll 1$
in (\ref{two-pointpert}) and write a  simple representation for the
$x$-dependent couplings as
\be
\label{cutoff}
\hat{\phi}_0(\bar{x})=L^{-2}\int\rmd^4\bar{y}\,\delta^{4}(
\bar{x}-\bar{y})\,. 
\ee
Then, using  translation invariance one can show that  only the unit
term of the 
OPE coefficient ${\cal C}_{{\cal O}^2}$ survives the UV limit
$L\rightarrow 0$ of (\ref{two-pointpert}). Since now 
(\ref{two-pointfinal}) is to be interpreted as the deformation of the
CFT by a coupling of the form $\,g\!\int\!{\cal O}^2_{ren}$, the
$O(\lambda^2)$ terms could be absorbed in a renormalization of the
operator ${\cal O}^2$ as
\be
\label{O2renorm}
{\cal O}^2(\bar{x}) =
[1-\frac{\lambda^2\gamma_*}{2} 
\ln(\bar{x}^2\mu^2)-\lambda^2b_*]{\cal O}^2_{ren}(\bar{x})\,,
\ee
and then (\ref{two-pointfinal}) becomes by virtue of (\ref{anom2})
\be
\label{two-pointfinal1}
\langle{\cal O}(\bar{x}_1){\cal O}(\bar{x}_2)\rangle_{R}' =
\langle{\cal O}(\bar{x}_1){\cal O}(\bar{x}_2)\rangle_R
+\left. \frac{G_{{\cal O}^2}}{C_{{\cal 
    O}^2}}\right|_0\langle\Theta\rangle \int \rmd^4\bar{x}\langle {\cal
  O}^2_{ren}(\bar{x}){\cal    
 O}(\bar{x}_1) {\cal O}(\bar{x}_2)\rangle_R+\cdots\,.
\ee

\section{Conclusions and outlook}

For the simple model of a scalar field in AdS$_5$ with cubic
self-interaction we showed that deformations of the
boundary CFT by scalar operators which appear in the OPE of the basic
operators ${\cal O}$  can be 
incorporated in the context of AdS/CFT by an appropriate  adjustment  of
the boundary data 
$\phi_0$. From the point of view of the bulk theory such an adjustment
should correspond to a change in the boundary condition for the bulk
fields. From the point of view of the boundary theory such an adjustment
is a reparametrization of the $x$-dependent coupling associated with
the boundary operator ${\cal O}$. It would be nice to study the
relation between our 
reparametrization formulae (\ref{cond1})-(\ref{phi1}) and the
``boundary equations of motion'' of \cite{BSS}. 
Our results can be generalized to
the case when the OPE contains many more relevant and marginal operators.

Essential role in our calculations played the OPE of the boundary
fields ${\cal O}$. In the ${\cal N}=4$ SYM case, the generalization of
our results should provide a concrete method to study multi-trace
deformations of the boundary CFT via OPEs of single-trace
operators. Moreover, it is conceivable that our method could also
be useful in studies of deformations of the (2,0) tensor multiplet in
$d=6$ at strong coupling, despite the fact that in this case there
exists no notion of ``multi-trace'' operators. In particular, in the
${\cal N}=4$ SYM case one knows
a list of double-trace operators in various representations of the
$SU(4)$ symmetry group that have naive dimension 4 and could in
principle be incorporated by our method. Furthermore, their three-point
function couplings and two-point function normalization constants are
known. One has
to distinguish between the various operators that are deforming the
boundary CFT. For example, the
double-trace operator in the $\bf{[1]}$ of $SU(4)$,
denoted $\bf{{\cal O}_1}$ in \cite{AFP1}, acquires anomalous
dimensions of order $1/N^2$ at strong coupling. Therefore, to
incorporate such a deformation one needs to know the  anomalous
dimension of the operator as well as the correction to its
coupling. On the other 
hand, the double-trace operator in the 
$\bf{[20]}$ of $SU(4)$, denoted $\bf{{\cal O}_{20}}$ in \cite{AFP1},
has protected 
dimension and can be incorporated more easily. It is highly probable
that both operators above
break the conformal invariance of the theory, although the case of
$\bf{{\cal O}_{20}}$ deserves further study.

In explicit AdS/CFT calculations, bulk actions such as
(\ref{action}) do not in general include arbitrary parameters since
the relative coefficients of all terms are 
fixed by SUGRA. Nevertheless, deforming the boundary CFT by a double-trace
operator entails the introduction of an arbitrary ``bare'' coupling
constant. As shown in (\ref{two-pointfinal1}), for marginal
deformations  this is nothing but the
integrated conformal anomaly. In the simple model studied here,
(\ref{anom}) is the only conformal 
anomaly in the boundary CFT, however in explicit SUGRA calculations we
expect that the gravitational anomaly will also play a role as a bare
coupling for double-trace marginal deformations. The
implications of such a result for a better understanding of the
double-trace operators in terms of string theory, either as non-local
string couplings or otherwise, is an interesting question.

\subsection*{Acknowledgements}

I would like to thank G. Arutyunov for very interesting discussions as
well as a critical reading of the manuscript. I would also like to
thank W. M\"uck for his interest and his constructive critisism.

\section*{Appendix}
We give for completeness the explicit representation for the
bulk-to-bulk propagator needed for the calculations of the boundary
effective action.
\bea
\label{G}
G(x,y) &=&
-c_{\Delta}\xi^{-\Delta}{}_2F_1(\frac{\Delta}{2}+\frac{1}{2},
\frac{\Delta}{2};\Delta-1;\xi^{-2})\nonumber \\
\label{xi}
\xi^2 &=& \frac{r^2+r'^2+(\bar{x}-\bar{y})^2}{2rr'} \,, \,\,\,\,\,\,
  c_{\Delta} = \frac{\Gamma(\Delta)}{2^{\Delta +1}\pi^2\Gamma(\Delta
  -1)} \nonumber \\
\label{F}
F(x,y)&=& \int\frac{\rmd^4\bar{p}}{(2\pi)^4}e^{-{\rm
    i}\bar{p}(\bar{x}-\bar{y})} (rr')^{2}{\rm
    K}_{\a}(|\bar{p}|r){\rm K}_{\a}(|\bar{p}|r')\frac{{\rm
    I}_{\a}(|\bar{p}|\epsilon)}{{\rm K}_{\a}(|\bar{p}|\epsilon)}
    \nonumber \\
\label{dG}
\partial_{r}G_{\epsilon}(x,y)|_{r=\epsilon} &=&
-\epsilon^{3}K(\epsilon;\bar{x}, r';\bar{y})\nonumber 
\eea

The explicit expression we use for the correlators appearing in
(\ref{Sren}) are (for non-coincident points and general $d$) \cite{HPR}
\bea
\label{corrs}
\langle {\cal O}(\bar{x}_1){\cal O}(\bar{x}_2)\rangle &=& C_{\cal
  O}\frac{1}{ \bar{x}^{2\Delta}_{12}}\nonumber \\
\langle {\cal O}(\bar{x}_1){\cal O}(\bar{x}_2){\cal O}(\bar{x}_3)
  \rangle &=& \frac{1}{4\pi^d}\frac{\Gamma^3(\frac{\Delta}{2})
  \Gamma(\frac{3\Delta}{2}
  -\frac{d}{2})}{\Gamma^3(\Delta-\frac{d}{2})}
  \frac{1}{(\bar{x}^2_{12}
  \bar{x}^2_{13}\bar{x}^2_{23})^{\frac{\Delta}{2}}} \nonumber \\
\langle {\cal O}(\bar{x}_1){\cal O}(\bar{x}_2){\cal O}(\bar{x}_3)
  {\cal O}(\bar{x}_4)\rangle &=& \frac{1}{\lambda^2}\langle {\cal
  O}(\bar{x}_1){\cal 
  O}(\bar{x}_2){\cal O}(\bar{x}_3) 
  {\cal O}(\bar{x}_4)\rangle_{disc} \nonumber \\
&& +\langle {\cal O}(\bar{x}_1){\cal
  O}(\bar{x}_2){\cal O}(\bar{x}_3) 
  {\cal O}(\bar{x}_4)\rangle_{conn} \nonumber \\
\langle {\cal O}(\bar{x}_1){\cal
  O}(\bar{x}_2){\cal O}(\bar{x}_3) 
  {\cal O}(\bar{x}_4)\rangle_{disc} &=& \langle {\cal O}(\bar{x}_1){\cal
  O}(\bar{x}_2)\rangle\langle {\cal O}(\bar{x}_3) 
  {\cal O}(\bar{x}_4)\rangle +(\bar{x}_2 \leftrightarrow \bar{x}_3)
  +(\bar{x}_2 \leftrightarrow \bar{x}_4) \nonumber \\
\langle {\cal O}(\bar{x}_1){\cal
  O}(\bar{x}_2){\cal O}(\bar{x}_3) 
  {\cal O}(\bar{x}_4)\rangle_{conn}  &=&   
-\int_{0}^{\infty}\frac{\rmd 
  r\rmd r'}{(rr')^{d+1}}\int\rmd^d\bar{x}\rmd^d\bar{y}\Bigl[
  \hat{K}(r;\bar{x},,\bar{x}_1) \hat{K}(r;\bar{x},,\bar{x}_2)
  G(x,y) \nonumber \\
&&\hspace{2cm} \hat{K}(r';\bar{y},\bar{x}_3)
  \hat{K}(r';\bar{y},\bar{x}_4)\Bigl]+(\bar{x}_2 \leftrightarrow \bar{x}_3)
  +(\bar{x}_2 \leftrightarrow \bar{x}_4) \nonumber \\
\hat{K}(r';\bar{y},\bar{x}) &=& C_{\Delta}\left[\frac{r'}{r'^2
  +(\bar{y}-\bar{x})^2} \right]^{\Delta}\nonumber 
\eea

\end{document}